\documentclass[11pt]{article}

 \usepackage{amsmath}
 \usepackage{graphicx}
 \usepackage{amssymb}
 \usepackage{amsfonts}
 \usepackage{latexsym}

\renewcommand{\baselinestretch}{1.1}
\renewcommand{\vec}[1]{{\bf #1}}

\def\beq{\begin{eqnarray}}
\def\eeq{\end{eqnarray}}

\def\al{\alpha}
\def\be{\beta}

\def\ga{\gamma}
\def\de{\delta}

\def\la{\lambda}
\def\na{\nabla}
\def\pa{\partial}

\def\si{\sigma}

\def\Ga{\Gamma}

\def\Si{\Sigma}

\begin{document}
\begin{center}
{\large\bf The Exact Foldy-Wouthuysen Transformation for a Dirac Theory with the complete set of CPT-Lorentz 
violating terms}
\vskip 6mm
\textbf{Bruno Gon\c calves\footnote{
E-mail: bruno.goncalves@ifsudestemg.edu.br},
\
M\' ario M. Dias J\' unior\footnote{
E-mail: mardiasjr@yahoo.com.br}
\ and \
Baltazar J. Ribeiro\footnote{
E-mail: baltazarjonas@nepomuceno.cefetmg.br}
}
\vskip 4mm
$^{1,2}$Instituto Federal de Educa\c c\~ ao, Ci\^ encia e Tecnologia Sudeste de Minas Gerais \\
\centering{IF Sudeste MG, 36080-001, Juiz de Fora - MG, Brazil}\\
\vspace{0.4cm}
$^{3}$Centro Federal de Educa\c c\~ ao Tecnol\' ogica \\
CEFET, 37.250-000, Nepomuceno - MG, Brazil

\end{center}
\vskip 4mm
\begin{quotation}



\begin{abstract}
The exact Foldy-Wouthuysen transformation is performed in order to study the Dirac
field interacting with many possible external fields  
associated with CPT-Lorentz violation. We also derived the calculation of equations of motion 
as well as the generalized Lorentz force corrected by the mentioned external fields.
The main point is the interaction between the Dirac particle and the terms that have 
the multiplication of electromagnetic field and the terms that break CPT-Lorentz.
Finally, with the transformed Hamiltonian we were able to write an expression for the 
bound state of the theory and analyze it in the atomic experiments context. This result is 
an analytical expression that gives the possibility of the weakness of CPT/Lorentz terms to
be compensated by the presence of a strong magnetic field. 
\end{abstract}


\vspace{0.2cm}
\end{quotation}
\vskip 4mm

\section{Introduction}

\hspace{0.65cm}The study of the possibility of the violation of CPT-Lorentz symmetries has received significant attention of the 
scientific community in the last years. For example, the CPT-Lorentz symmetries violation
caused by terms in the action of quantum fields was considered in \cite{costelchy1}. There are also great efforts 
about studding weakly broken CPT-Lorentz symmetries not only in the mathematical point of view. 
In Ref. \cite{costelechy2}, it is possible to see that some physical consequences may arise from
the assumption that these symmetries can be in fact introduced into the action of the theory. 
In this context, this subject becomes important not only for theoretical physics. There is a growing expectation 
about experimental evidences of such violation. In Ref. \cite{colladay} one can find an overview of 
all the possibilities of breaking CPT-Lorentz symmetries. There are many other recent Refs. 
(\cite{baet1,CFMS}) that treat the possibility of CPT-Lorentz symmetry breaking in a more 
phenomenological point of view. But, in fact, the main part of the study of scientific community remains on 
the search for the correct theory that could give the right interpretation for this symmetry breaking.  

It is possible to find many references in 
the literature that support the evidence of this form 
of symmetry breaking. The parameters for experimental verification scenario are completely related to the 
consistency at the quantum level of the theory. For this reason the renormalization 
structure of theories with broken symmetries is an important feature to be analyzed 
(see, e.g., Ref.\cite{shapiro1} for a brief and qualitative review of the renormalization
properties of quantum field theories in curved spacetime in the presence
of CPT-Lorentz violating terms). 

A possible connection between theories with CPT-Lorentz
violating terms and Cosmology (for baryogenesis considerations see \cite{bertolami}) 
worth to be analyzed, especially in the remarkable 
scenario of cosmic microwave radiation. In Ref. \cite{conf} there is an overview 
about the possibility of the Lorentz symmetry breaking manifestation. A possible candidate to produce the 
violation is the torsion field (except for unitarity problem). It is very interesting to see 
\cite{shapirorep,costelechyprd} that this field could be generated from the symmetry breaking. These papers 
are showing the connection between the mathematical symmetry violation and some physical theory. 
It is also known that the torsion field, if it really exists, is very weak. But the main 
point here, is to achieve an interpretation to each of the CPT-Lorentz terms in such a way that it would be possible 
to search for some experimental result to support these theoretical results. The interaction between the spacetime 
torsion and the spinor field (and the spinning particle) is described in \cite{data,audret,hayashi,rumpf}. 
In fact, not only torsion, but all the appropriated CPT-Lorentz violating terms would be one of the possible 
candidates  
to play the role to describe the known anisotropy related to cosmological perturbations. Another example is shown in 
\cite{tiba} were the photon contribution to the divergences and conformal anomaly in the theory 
with CPT-Lorentz violating terms has been evaluated.  

However, if we focus our attention on the torsion field (that has direct connection to CPT-Lorentz violation), 
in order to obtain a better interpretation about physical properties of the  
Dirac equation, it is important to consider the nonrelativistic approximation of the theory. 
In the Refs. \cite{sabbata,bagrov,ryder}, the 
Foldy-Wouthuysen transformation (FWT) has been 
obtained for the fermion field coupled
to the combined electromagnetic and torsion fields. The FW method is a 
direct calculation that allows to extract physical information of the 
Dirac action (see e.g. \cite {JN2, Ob}). It is one of the methods to extract physical information from the 
Dirac Hamiltonian that is not diagonal from the beginning. That is the case when the CPT-Lorentz terms 
are taken into account.

Furthermore, it is possible to study the cases when not only the torsion field is considered.
We can now think of which would be the appropriate approach to get physical information
from a Hamiltonian that contains the other different terms that break CPT-Lorentz. The first
point would be to choose the correct Hamiltonian, once none of the terms is experimentally
verified. It can not be constructed using some empirical insights. But, the question here is a little bit more complicated. 
After the analytical Hamiltonian is presented, performing the FWT is not straightforward. The calculations 
may become cumbersome and, in the practical point of view, the analysis of the result could be lost 
into a very complex expression for the transformed equation. Then, this method, that was used before, 
when the unique term of interest was the torsion, seems to not work when the intention is to 
perform the calculus for more CPT-Lorentz breaking terms.  

In this paper, we discuss the method to derive the physical analysis from the initial Hamiltonian. The 
EFWT is the more economic (in the algebra) and the more reliable approach to give the information 
in which we are interested. Another advantage that this transformation has is the fact that it allows us 
to use some specific notations that make all the steps, during the calculations, very clear. Then, in the 
end, the transformed equations have results that can be easily compared to the ones known from the literature.

In Ref. \cite{shapiro}, there is an algorithm that shows how to construct a general Hamiltonian describing 
the $80$ new cases of CPT-Lorentz violating terms in the Dirac equation that can be analyzed with the EFWT. 
The focus of the present work is precisely the development of EFWT, considering the whole situation presented in the last 
related work as well as the calculation of equations of motion. So, here we are going to treat the torsion as a particular 
case of the equations presented here, as well as each other terms in the complete Hamiltonian. Our aim is to get the 
most general result that 
can be written to perform the physical analysis for the CPT-Lorentz violating terms using the method we believe that is 
the most complete one to perform this kind of study, that is the EFWT. We also explore the possibility of combining such equations 
in order to obtain an expression to describe the dynamic of the particle.

Although FWT \cite{fw} provides, in general, more detailed information about
the nonrelativistic approximation, there is a considerable advantage in 
the construction of the EFWT \cite{case,eriksen,nikitin,tiomo,obukhov,silenko}. 
Using the Exact transformation, the risk of missing 
some important terms is lower \footnote{See \cite{obukhov} for the spinor field in the weak gravitational field case.}.  
However, the possibility of performing exact transformation depends on the existence of the so called 
involution operator on the external classical fields and, from the mathematical point of view, 
EFWT is more complex \cite{eriksen,nikitin}. 


The paper is organized as follows. In the next section we present a brief summarize about EFWT. 
Coming up we consider the $80$ 
cases mentioned above of CPT-Lorentz violating terms in the Dirac equation and perform EFWT. After that, the equations of motion are 
presented as a section and a more general Lorentz force is obtained. In the section 5 we present an equation for the bound state 
of this theory. Such equation is analyzed in the context of the CPT-Lorentz atomic experiments. In section 6 we draw our conclusions. Throughout 
the paper we use Greek letters for the indices which
run from 0 to 3. Latin indices are used for the space coordinates and run from 1 to 3.

\section{Exact Foldy-Wouthuysen transformation}

\hspace{0.65cm}We present in this section a brief review about EFWT \cite{obukhov}. We begin with the 
representation of even and odd operators in the FW context 
\begin{equation} 
M_{(EVEN)}=\frac{1}{2}(M+\be M \be)\hspace{0.8cm}\mbox{and}\hspace{0.8cm}
M_{(ODD)}=\frac{1}{2}(M-\be M \be)\label{evenodd}\,.
\end{equation}

The spinor transforms according to the following equations
\begin{equation}
\psi^{tr}=U\psi,\hspace{1cm}\psi=U^*\psi^{tr},
\end{equation}
where $U$ is an unitary operator. After leading the last equation to the relation
$i\partial\psi/\partial t=H\psi$ we have
\begin{equation}
{\cal H}^{tr}=U{\cal H}U^*-iU\dot{U}^*, \label{htr}
\end{equation}
where we have considered $i\partial\psi^{tr}/\partial t={\cal H}^{tr}\psi^{tr}$. We focus on time
independent external fields, then the last term in the equation (\ref{htr}) is zero. After considering
the commutation relation $[\be,{\cal H}^{tr}]=0$, one can show that $[U^*\be U,{\cal H}]=0$ and a possible choice to play the hole 
of the quantity $U^*\be U$ is
\begin{equation}
U^*\be U=\frac{{\cal H}}{\sqrt{{\cal H}^2}}=\la\,.\label{bilbo}
\end{equation}
In this case $\sqrt{{\cal H}^2}$ must be understood as a notation. The calculation of ${\cal H}^2$ is performed in the coordinate representation and
as a next step it is necessary to write ${\cal H}^2$ in the momentum representation in order to derive its squared root. 
In the last equation, the quantity $\la$ is a Hermitian and unitary operator $\la^2=\la^{\dagger}\la=1$. 
We assume that the Hamiltonian is well defined and it does not possess zero eigenvalues \cite{eriksen,nikitin}.
It is remarkable to say that to perform EFWT, the Hamiltonian must obey the following relation
\begin{equation} 
J{\cal H}+{\cal H}J=0\,,\label{gyp}
\end{equation}
where
\begin{equation}
J=i\ga^5\be
\end{equation}
is Hermitian and unitary. It is important to note that this explicit form of involution operator 
is the one used in \cite{shapiro} to establish the criteria to perform the EFWT for CPT-Lorentz terms. 
Another useful information about $J$ is the relation 
$J\be+\be J=0$. The EFWT is performed if we consider the relation $U=U_2\, U_1$ where
\begin{equation}
U_1=\frac{1}{\sqrt{2}}(1+J\la)\hspace{0.8cm}\mbox{and}\hspace{0.8cm}
U_2=\frac{1}{\sqrt{2}}(1+\be J)\,.\label{pid}
\end{equation}
Consequently 
\begin{equation} 
U_{1}\la U_{1}^{\dagger}=J\hspace{0.8cm}\mbox{and}\hspace{0.8cm}
U_{2}\la U_{2}^{\dagger}=\be\,.
\end{equation}

It is not difficult to check that the equations (\ref{pid}) satisfy the relation $U\la U^{\dagger}=\be$.
From the equation (\ref{bilbo}) it is easy to see that
\begin{equation}
\be=U\la U^*=\frac{{\cal H}^{tr}}{\sqrt{({\cal H}^{tr})^2}}
\end{equation}
and the quantity ${\cal H}^{tr}$ is given by ${\cal H}^{tr}=\be\sqrt{({\cal H}^{tr})^2}$ which is an even Hamiltonian. 
Now and so on we denote the terms with ``tr'' index as the transformed ones 
and such terms belong to the final transformed Hamiltonian.

In the practical
sense, performing all the steps to get ${\cal H}^{tr}$ is necessary to know the operator $U$ as a solution of the equation 
(\ref{bilbo}). Let us consider $U=\sqrt{\be\la}$ as a solution. The operator $U$ has the property $U\be=U^*\be$. 
Now, from equation (\ref{gyp}) we have $J\sqrt{{\cal H}^2}=\sqrt{{\cal H}^2}J$ and $J\la+\la J=0$. 
These last relations imply that $U_1 {\cal H} U{_1}^{\dagger}=J\sqrt{{\cal H}^2}$ and we finally 
have the transformed Hamiltonian 
\begin{equation}
{\cal H}^{tr}=U{\cal H}U^*=\be[\sqrt{{\cal H}^2}]_{(EVEN)}+J[\sqrt{{\cal H}^2}]_{(ODD)}, \label{trunf}
\end{equation}
where even and odd operators are defined in equation (\ref{evenodd}).  
The quantities $\sqrt{{\cal H}^2}_{(EVEN)}$ and $\sqrt{{\cal H}^2}_{(ODD)}$ represent the even and odd squared Hamiltonian respectively. 
If $[{\cal H}^2,\be]$=0 we have an even term, and if $\{{\cal H}^2,\be\}=0$ an odd one.
The equation (\ref{trunf}) represents the EFWT. 
As a second step one should 
calculate ${\cal H}^2$ 
with the even and odd correct identification and then multiply them by either $\be$ or $J$. 
The operation $\sqrt{{\cal H}^2}$ can be 
considered in different approaches. A simpler option is to write ${\cal H}={\cal H}_0+{\cal H}_{int}$. The coupling 
constants of the interaction terms are present in ${\cal H}_{int}$. That is, in fact, the main difference 
between the usual FWT and the EFWT. The freedom in the choice of the expansion parameter may give us results 
with more familiar analytical form for the final expressions. This fact, in general, makes the physical 
analysis easier to perform. The free Hamiltonian 
is denoted by ${\cal H}_0$ and the interaction one by ${\cal H}_{int}$.
Both terms in the equation (\ref{trunf}) are even terms and for this reason 
the quantity ${\cal H}^{tr}$ does not mix upper spinor components with lower ones.

\section{Exact transformation with CPT}
\hspace{0.65cm}We shall deal with the action of a Dirac fermion theory, including CPT-Lorentz symmetry breaking terms
\begin{equation}
S=\int d^4x \sqrt{-g}\Big\{\frac{i}{2}\bar{\Psi}\,\Ga^{\mu}D_{\mu}\Psi-\frac{i}{2}D_{\mu}^{\star}
\bar{\Psi}\,\Ga^{\mu}\Psi-\bar{\Psi}M\Psi\Big\}\,.\label{action}
\end{equation}

The CPT-Lorentz symmetry breaking terms can be written in the following way \cite{colladay}
\begin{equation}
D_{\mu}=\na_{\mu}-ieA_{\mu};\hspace{1cm}D^{\star}_{\mu}=\na_{\mu}+ieA_{\mu};
\hspace{1cm}\Ga^{\mu}=\ga^{\mu}+\Ga^{\mu}_{1};\hspace{1cm}M=m+M_1\,.
\end{equation}

The quantities $\Ga^{\nu}_{1}$ and $M_1$ can be written in terms of the CPT-Lorentz 
violating parameters\footnote{The authors discuss in ref. \cite{kostelecky2,jackiw2} some aspects about origin of these terms.}
 $a_{\mu}$, $b_{\mu}$, $m_{5}$, $c^{\mu\nu}$, $f^{\mu}$, $e^{\mu}$, $g^{\mu\nu\la}$, $H_{\mu\nu}$, 
in the following way
\begin{equation}
\Ga^{\nu}_{1}=c^{\mu\nu}\ga_{\mu}+d^{\mu\nu}\ga_5\ga_{\mu}+e^{\nu}+if^{\nu}\ga_5
+\frac{1}{2}g^{\la\mu\nu}\si_{\la\mu}
\end{equation}
and
\begin{equation}
M_1=a_{\mu}\ga^{\mu}+b_{\mu}\ga^5\ga^{\mu}+im_5\ga^5+\frac{1}{2}H^{\mu\nu}\si^{\mu\nu}\,.
\end{equation}

We adopt notations as described in \cite{bjorken} for Dirac Matrices and the 
useful notations for $P_i$, used in \cite{shapiro}
\beq
{P}^0_{\nu} = (0,P_i)\,\,,
\qquad 
\overline{P}_{\nu} = {P}^0_{\nu} - e A_{\nu}
\qquad \mbox{and} \qquad
P_{\nu}^* = \overline{P}_{\nu} - \frac{i}{2}\na_\nu\,.
\eeq
Under such definitions, equation (\ref{action}) takes the form
\begin{equation}
S=\int d^4x \bar{\Psi}\big\{i\Ga^{\mu}D_{\mu}+\frac{i}{2}(\na_{\mu}\Ga^{\mu})-M\big\}\,.
\end{equation}

After such considerations, taking the equations of motion for $\psi$ in the Schrodinger 
form $i\partial_{t}\psi={\cal H}\psi$, one can check that
\begin{equation}
i\na_0\psi=\big\{\ga_0-\ga_0(c^{\mu0}\ga_{\mu}+d^{\mu0}\ga_5\ga_{\mu}+e^0+if^0\ga_5
+\frac{1}{2}g^{\la\mu0}\si_{\la\mu})\ga_0\big\}\times[M+(P^{*}_{\nu}\Ga^{\nu})]\Psi\,.
\label{gali}
\end{equation}

In order to perform the EFWT, it is necessary that the Hamiltonian admits the 
involution operator \cite{case,eriksen,nikitin,obukhov}. 
The most general form of the equation (\ref{gali}) that admits the involution 
operator is presented in Ref. \cite{shapiro}. In such work the authors present a complete table which 
contains the 80 cases of CPT-Lorentz violating terms in the modified 
Dirac equation that admits EFWT. For the sake of completeness, we consider in the present
work the complete Hamiltonian with CPT-Lorentz breaking terms of Dirac Theories admitting EFWT
\begin{eqnarray}
{\cal H}&=&m\Big(\ga^0-\ga^0c_{00}+i\ga^5f_0 +\ga^i\ga^5d_{i0}+\frac{i}{2}\al^{i}g_{i00}\Big)
+\bar{P}_l\Big(\ga^0e^l+\al_{i}c^{il}-\ga^5d^{0l}+ \al^{l}\nonumber\\
&+&\frac{1}{2}\ga^{0}\si^{ij}\,g_{ijl} 
-\al^{l}c^{00}+i\ga^5\ga^l f_0+\ga^i\ga^5\ga^ld_{i0}+\frac{i}{2}\al^i \ga^l g_{i00}\Big)\nonumber\\
&+&\al^l a_l -\ga^5b_0+\frac{1}{2}\ga^0 \si^{ij}H_{ij}\,.\label{frst}
\end{eqnarray}

Interaction terms of second order are very small and for this reason they can be neglected in this last 
equation and so on. We calculate the squared Hamiltonian as a first step to get exact transformation. 
We adopt the quantity $\bar{{\cal H}}^{2}$ to represent the squared Hamiltonian.
It is convenient to write the quantity $\bar{{\cal H}}^{2}$ in the form
\begin{equation}
\bar{{\cal H}}^{2}=(1+\bar{A})[(\delta_{ij}+B_{ij})\bar{P}^i
+\bar{C}_j]^2 +\bar{D}+m^2\,,\label{sid}
\end{equation}
where we define the following notations
\begin{eqnarray}
\bar{A}&=&-4d_{i0}\ga^{5}\al^{i}-4if_0 \ga^0\ga^5+ig_{i00}\ga^0 \al^i\,,\nonumber\\
\bar{B}_{ij}&=&\frac{1}{2}\Big[-2c_{ij}-2\ga^5\al^i d_{0j}-2\ga^5\al^i d_{j0} 
- 2\ga^{0}\ga^{5}\epsilon^{lmi}g_{lmj}+ 2i\ga^{0}\ga^{5}g_{tij}\Sigma^{t}+\frac{i}{3}g_{i00}\ga^{0}\al_{j}\Big]\,,\nonumber\\
\bar{C}_{j}&=&\frac{1}{2}\Big[2m\,e_{j}+m\si^{lm}g_{lmj}-2img_{j00}
-4m\ga^{0}\ga^{5}d_{j0}  -2a_j-2\ga^{5}\al^{j}b_0-2H_{lm}\epsilon^{lmj}\ga^{0}\ga^{5}\nonumber\\
&-&m g_{l00}\epsilon^{ljk}\Sigma_{k}-2im \ga^{0}\ga^{5}d_{m0}\epsilon^{jmk}\Sigma_{k}
+2i \ga^{0}\ga^{5}H_{mj}\Sigma^{m}\Big]\,,\nonumber\\
\bar{D}&=&-2m^2c_{00}-2m^2\ga^{5}\al^{i}d_{i0}+ m \si^{ij} H_{ij}\nonumber\\
&+&\Big(1-2id_{i0}\ga^{5}\al^i-2if_0 \ga^{0}\ga^{5}
+\frac{i}{3}g_{i00}\ga^{5}\al^i\Big)\frac{i\hbar e}{mc}\Si_{k}B^{k}\,.\label{dus}
\end{eqnarray}
These notations are used to simplify the algebra and make the interpretation 
of the results easier to perform. In this form, it is 
possible, for example, to direct identify which terms are linked 
with the kinetic part of the Hamiltonian.
 
In accordance to EFWT transformation, we should take squared root of even terms in equation 
(\ref{dus}) and multiply them by $\be$. On the other hand, the squared root of odd terms should be multiplied 
by $J=i\ga^5\be$. Moreover, such procedure is not quite trivial in the situation when there are 
many odd terms. We present a new point of view to perform the exact transformation. Note that 
it is possible to show that the equation ({\ref{trunf}) 
is completely equivalent to
\begin{equation}
{\cal H}^{tr}=J\frac{1}{2}(\sqrt{{\cal H}^2}-\be \sqrt{{\cal H}^2}\be)
+\be\frac{1}{2}(\sqrt{{\cal H}^2}+\be \sqrt{{\cal H}^2}\be)\,.\label{siga}
\end{equation}
This last relation allows us to perform the EFWT where many odd terms are present. 
According to equation (\ref{siga}), it is possible to take as a first step, the squared root of the complete Hamiltonian
and as a second step, identify and separate even and odd terms.

In order to take the squared root we shall consider that $m^2\gg \bar{{\cal H}}^2$, in the equation 
(\ref{sid}), so well as the expansion. 
These considerations allow us to assert that the following equation
\begin{equation}
{\cal H}^2=m^2\Big(1+\frac{\bar{{\cal H}}^2}{m^2}\Big)
\end{equation}
can be written as 
\begin{equation}
\sqrt{{\cal H}^2}=m\,\Big(1+\frac{\bar{{\cal H}}^2}{2m^2}\Big)\,,
\end{equation}
where $\bar{{\cal H}}^2$ is given by relation (\ref{sid}). After some algebra we have the following equation 
for the transformed Hamiltonian
\begin{equation}
{\cal H}^{tr}=\be m+\frac{1}{2m}(1+A^{tr})[(\delta_{ij}+B_{ij}^{tr})\bar{P}^i
+C_j^{tr}]^2 +D^{tr}\,,\label{sedd}
\end{equation}
where
\begin{eqnarray}
A^{tr}&=&-4\be\Sigma^{i}d_{i0}+4f_0 +\Sigma^{i}g_{i00}\,,\nonumber\\
\bar{B}_{ij}^{tr}&=&\frac{1}{2}\Big[-2\be c_{ij}+2\be \Sigma^i d_{0j}+2\be \Sigma^i d_{j0} 
- 2i \epsilon^{lmi}g_{lmj} -2 g_{tij}\Sigma^{t}+\frac{1}{3}g_{i00}\Sigma_{j}\Big]\,,\nonumber\\
\bar{C}_{j}^{tr}&=&\frac{1}{2}\Big[2m \be\,e_{j}+m\be g_{lmj}\epsilon^{lmk}\Sigma_{k}
-4im d_{j0}  -2\be a_j+2\be\Sigma^{j}b_0-2i\epsilon^{lmj}H_{lm}\nonumber\\
&-&m \be \epsilon^{ljk}\Sigma_{k}g_{l00}+2m \epsilon^{jmk}\Sigma_{k}d_{m0}
-2im\be g_{j00} -2\Sigma^{m}H_{mj}\Big]\,,\nonumber\\
\bar{D}^{tr}&=&-2m^2\be c_{00}+2m^2\be \Sigma^{i}d_{i0}+ m \be \epsilon^{ijk}\Sigma_{k} H_{ij}\nonumber\\
&+&\Big[\be(1+2i\Sigma^{i} d_{i0})+2f_0 
+\frac{1}{3}\Sigma^{i}g_{i00}\Big]\frac{i\hbar e}{mc}\Si_{k}B^{k}\,.\label{dus}
\end{eqnarray}

The equation (\ref{sedd}) presents a well known structure. According to this equation, 
the first term corresponds to rest energy. The second one represents the kinetic term. 
It is a term of the kind $\vec{P}-e\,\vec{A}$ and one can imagine the quantity $C_{j}^{tr}$ 
(in the situation where $B_{ij}^{tr}=0$), as being an analogous term 
of a gauge transformation to $\bar{P}^{i}$. The quantity $1+A^{tr}$ can be seen as a  correction to the general 
form of the kinetic energy. The last term in the equation (\ref{sedd}) corresponds to a  pure external interaction.

\section{Equations of motion}
\hspace{0.65cm}We perform in this section the calculations of equations of motion. 
We also explore the possibility of combining such equations 
in order to obtain an expression to describe the dynamic of the particle as a result. The Lorentz force 
corrected by CPT-Lorentz violating terms is obtained. Let us begin by taking into account the two components spinor 
\begin{equation}
\psi=\left(
\begin{array}{ccc}
\phi \\
\chi  \\
\end{array}
\right) \,\exp^{-imt}
\end{equation}
and write the Dirac equation in the Schr\" odinger form  $i\partial_{t}\psi={\cal H}\psi$. After some algebra, one can get the 
following Hamiltonian to $\phi$
\begin{equation}
{\cal H}= \frac{1}{2m}\big\{(1+A)[\big(\de_{ij}
+B_{ij}\big)\bar{P}^i+C_{j}]^2+D\big\}\,,\label{nos}
\end{equation}
where
\begin{eqnarray}
A&=&-4\sigma^{i}d_{i0}+4f_0 +\sigma^{i}g_{i00}\,,\nonumber\\
B_{ij}&=&-c_{ij}+\sigma^i d_{0j}+\sigma^i d_{j0} 
-i \epsilon^{lmi}g_{lmj} - g_{tij}\sigma^{t}+\frac{1}{3}g_{i00}\sigma_{j}\,,\nonumber\\
C_{j}&=&m \,e_{j}+ \frac{1}{2}m g_{lmj}\epsilon^{lmk}\sigma_{k}
-2im d_{j0}  -a_j +\sigma^{j}b_0-i\epsilon^{lmj}H_{lm}\nonumber\\
&-&\frac{1}{2}m \epsilon^{ljk}\sigma_{k}g_{l00}+m \epsilon^{jmk}\sigma_{k}d_{m0}
-im g_{j00} -\sigma^{m}H_{mj}\,,\nonumber\\
{D}&=&-2m^2 c_{00}+2m^2 \sigma^{i}d_{i0}+ m \epsilon^{ijk}\sigma_{k} H_{ij}\nonumber\\
&+&\Big[1+2i\sigma^{i} d_{i0}+2f_0 
+\frac{1}{3}\sigma^{i}g_{i00}\Big]\frac{i\hbar e}{mc}\si_{k}B^{k}\,.\label{funkfunk}
\end{eqnarray}

In order to quantize Hamiltonian (\ref{nos}) and to write semi-classical equations of motion\footnote{After the calculus we 
make $\hbar\rightarrow 0$ in the same procedure adopted in \cite{buchbinder}.}
 let us consider the following relations
\begin{equation}
i\hbar \frac{d \hat{x}_i}{dt}=[\hat{x}_i,{\cal H}],\hspace{1cm}
i\hbar \frac{d \hat{p}_i}{dt}=[\hat{p}_i,{\cal H}]\label{quant}\hspace{1cm}
\mbox{and}\hspace{1cm}i\hbar\frac{d\,\hat{\si}_i}{dt}=[\hat{\si}_i,{\cal H}]\,.
\end{equation}
So we get
\begin{equation}
\frac{d \hat{x}_i}{dt}=\frac{1}{m}
\big(1+A\big)\big[\big(\de_{ij}+2B_{[ij]}\big)\bar{P}^j
+C_i\big]\,,\label{xis}
\end{equation}
\begin{eqnarray}
\frac{d \hat{p}_i}{dt}&=&\frac{1}{2m}
\Bigg\{-\frac{\partial A}{\partial x_i}
\big[(\de_{kj}+B_{kj})\bar{P}^k+C_{j}\big]^2\nonumber\\
&-&(1+A)\Big[2\Big(\frac{\partial }{\partial x_i}
[(\de^{lj}+B^{lj})\bar{P}_l+C_j]\Big)\big[(\de_{kj}+B_{kj})\bar{P}^k+C_j\big]
-\frac{\partial D}{\partial x^i}\Bigg\}\nonumber\\
&=&e\,v_j\frac{\pa A^j}{\pa x_i}\label{pe}
\end{eqnarray}
and
\begin{equation}
i\hbar \frac{d \hat{\si}_i}{dt}=\varepsilon_{ijk} \, R_j \, \si_k +C_{ij}\si^{j}\,,\label{si}
\end{equation}
where
\begin{eqnarray}
R_j&=&(-4imd_{j0}+img_{j00})v^2+\Big(2imd_{0l}\de_{jm}+2imd_{l0}\de_{jm}
-2img_{mlj}+\frac{2}{3}img_{l00}\de_{jm}\Big)v_{l}v_{m}\nonumber\\
&+&(2ib_0\de_{ij}+2iH_{ij})v^{i}+2im d_{j0}
-\frac{\hbar e}{m^2 c}(1+2f_0)B_j
\end{eqnarray}
and
\begin{equation}
C_{ij}=(2img_{kij}+img_{i00}\de_{jk}+2im d_{i0}\de_{jk}-img_{k00}\de_{ij}-2imd_{k0}\de_{ij})v^k+2i H_{ij}\,.
\end{equation}

Taking the derivative of equation (\ref{xis}) with respect to time one 
can get the generalized Lorentz force 
\begin{equation}
m\,\frac{dv_i}{dt}= \frac{d \la_{ij}}{dt}mv^{j}+\frac{d C_i}{dt}
+(\de_{ij}+\la_{ij})[-e \,\overrightarrow{E}\,+
e\,\overrightarrow{v} \times \overrightarrow{B}]^j \,,\label{princ}
\end{equation}
where
\begin{equation}
\la_{ij}=A\,\de_{ij}+2B_{[ij]}\label{lala}
\end{equation}
and the quantities $A$ and $B_{ij}$ are described by equations (\ref{funkfunk}). The equation 
(\ref{princ}) represents the Lorentz force corrected by CPT-Lorentz violating terms. 
It is worth noting that if the quantities $c_{00}$, $d_{i0}$, $c_{ij}$, $g_{lji}$ and $g_{i00}$ 
are null, one can get the Lorentz force as a particular case of equation (\ref{princ}).  
The first term in such equation represents a dragging term once it is proportional to the velocity. 
The second one is related to an external force and the last one is a correction to the the very well known quantity  
$e\,(\vec{E}+\vec{v}\times\vec{B})$.

\section{CPT tests, a perspective}

\hspace{0.65cm}We present in this section a brief perspective about experimental tests, tanking into 
account the EFWT whole scenario, including the 80 possible cases of CPT-Lorentz violation in the Dirac equation. 
As previously explained, we deal with the Hamiltonian of interaction between the Dirac spinor and external fields.

A consistent theoretical framework about CPT-Lorentz breaking tests 
(in particle and atomic systems) can be found in \cite{koste1,koste2,koste3}. 
Such framework incorporates CPT-Lorentz violation by using the so called standard model 
extension (SME) \cite{koste3} and the search for new signatures is possible. 
Nevertheless, this subject is very much extensive and a deep approach here would not be compatible with the 
context of this work. 

It is well known that Quantum Electrodynamics systems are extremely proliferous in the scope of CPT-Lorentz 
violation tests, since it is sensitive to very low energies. One can name some examples 
of experiments involving atomic physics experiments as Penning-Trap, Clock-Comparison, 
Torsion Pendulum, Hydrogen-Antihydrogen Experiments, 
Spin-Polarized Matter, Muon Experiments, among others (see refs. \cite{blumblum1,lane,blumblum2,blumblum3,blumblum4}, 
and references cited there in.). Each one of the experiments mentioned present a very specific bound state and 
the magnitude of such bounds makes possible to determine which kind of experiments should be performed \cite{koste5}.

A very natural question arises here. Is there a bound state associated to the EFWT for a Dirac 
theory related to the 80 new CPT-Lorentz violating terms? In order to answer this question, let us 
take into account the Lorentz violating potential $V$, 
which obeys the following relation \cite{lane}
\begin{equation}
V=-\tilde{b}_{j}\si_{j}\,,
\end{equation}
where $\si$ represents the spin matrices. After some algebra, one can 
write the corresponding bound in the following way\footnote{Note that the Lorentz violating potential comes, naturally, from 
the equation (\ref{funkfunk}).}
\begin{equation}
\tilde{b}_j=b_j-\frac{1}{2}\epsilon^{jlm} H_{lm} -m d_{j0} 
-\Big[1+2i\sigma^{i} d_{i0}+2f_0 
+\frac{1}{3}\sigma^{i}g_{i00}\Big]\frac{i\hbar e}{2 m^2 c}B_{j}\,.\label{guu}
\end{equation}
The bound presented in the last equation enables us to consider the possibility about getting an indication 
of possible atomic experiments \cite{koste5}. In fact, the magnitude of the magnetic field plays a crucial role 
in the determination of such experiments. However, direct analysis of the equation 
(\ref{guu}) 
to predict which is the most appropriate experiment is not straightforward.   

As an example of this procedure, we consider the following Hamiltonian obtained after 
Foldy-Wouthuysen transformation
\begin{equation} 
{\cal H}=m+\frac{p^2}{2m}+a_0-m c_{00}+\Big(-b_{j}+md_{j0}+\frac{1}{2}
\epsilon_{jkl}H_{kl}\Big)\si^{j}+\big[-a_j+m(c_{0j}+c_{j0})\big]\frac{p_j}{m}\,,\label{ramram}
\end{equation}
where all terms in the last equation where previously defined in section (3). This Hamiltonian  
presents the bound 
\begin{equation}
\tilde{b}_{j}=b_{j}-\frac{1}{2}\epsilon_{jkl}H_{kl}-m\,d_{j0}\label{bee}\,.
\end{equation}

Such bound is compatible with torsion pendulum experiments \cite{coste2000}. Clearly, the result 
presented in the equation (\ref{guu}) is more general than the one presented in the equation (\ref{bee}). 
If one considers the situation where the magnetic field is null, it is possible to get the bound state of the 
torsion pendulum experiment, described by (\ref{bee}) as a particular result. We also observe that 
the Hamiltonian described by the equation (\ref{ramram}) can be seen as a particular case of 
the equation (\ref{nos}). 

If we consider the same approach, that is, consider the only non-vanishing interaction terms to be 
$a_0\,,\,c_{00}\,,\,b_j\,,\,d_{j0}\,,\,H_{kl}$ and $c_{oj}$, from the equation (\ref{nos}), 
after some algebra, it is possible to obtain the same structure as the one observed on the equation (\ref{ramram}).
The terms $a_o$ and $b_j$ will not be present since they were not taken into account in initial Hamiltonian. 
The reason for this was the non acceptance of these terms to the EFWT criteria. Nevertheless, 
all the other quantities are present and have a more general form. In this sense it is possible 
to consider the equation (\ref{nos}) as a more complete case, despite the torsion field ($b_j$) and $a_j$ 
are not present. 

Another result that should be analyzed comes from the equation (\ref{princ}). This equation is related to the 
possibility of understanding the real particle behavior due to the interactions with the external fields. As an example, let us 
cite the interactions mixing terms between magnetic field $\vec{B}$ and the CPT-Lorentz terms of the kind $\la_{ij} B$. 

It is possible 
to imagine the case where the modulus of $\vec{B}$ is sufficiently high in order to compensate the weakness of the interactions. 
Possibility to measure such quantities in an indirect way (for example, using atomic and molecular physics) is 
contemplated if one considers the situation where a gas of electrons is present. In principle, it is possible to 
make a prediction of the motion generated by these new terms.

\section{Conclusions and discussions}
\label{Con}
\hspace{0.65cm}

The EFWT was here considered and performed in the context of 
all the 80 CPT-Lorentz violating terms, in the Dirac equation. 
The first result of the work is given by the equation (\ref{sedd}), which 
presents a well known structure and it is possible to identify in it terms like the rest, the kinetic and the potential energy. 
All of them corrected by the CPT-Lorentz violating terms.

It is worth noting that EFWT is not quite trivial, specifically in the presence of many odd terms.
Nevertheless, the equation (\ref{siga}) introduces a new way of performing the EFWT, such that it becomes 
easier to consider, in the situation when many odd terms are present. For this reason, the calculation procedure 
adopted here can be used as a guidance to the kind of situation described above.

Other results of the work are the semi-classical equations of motion for $\hat{x}_i$, $\hat{p}_i$ and $\hat{\si}_i$, 
given respectively by the equations (\ref{xis}), (\ref{pe}) and (\ref{si}). We have shown that it is 
possible to combine  equations of motion to get a generalized Lorentz force corrected by CPT-Lorentz violating terms, given 
by the equation (\ref{princ}). Such equation presents dragging term, external force and 
a factor of the kind $\de^{ij}+\la^{ij}$ 
correcting the quantity so much well known $e\,(\vec{E}+\vec{v}\times\vec{B})$. The quantity $\la^{ij}$ is given 
by the relation (\ref{lala}). As expected, one can get the usual Lorentz force by taking the limit in the absence of external fields in the equation ({\ref{princ}}).

In the last section we have highlighted that the bound state equation is important in order to get a better understanding 
about which kind of atomic experiments should be performed. In the context of the EFWT for a Dirac Theory with the
complete set of CPT-Lorentz violating terms, we have presented the bound state for this theory, given by the equation (\ref{guu}). 
Such equation allow us to conclude that the kind of atomic experiments that should be performed in this case, depends on the magnitude of the 
magnetic field. We also observe that if magnetic field is null, in the equation (\ref{guu}), on can recover the well known 
bound state, given by the equation (\ref{bee}), and the atomic experiment associated would be the pendulum experiments.

Another result from the last section is the possibility of understanding the particle behavior in the presence of 
interactions with the external fields. One can imagine interactions mixing terms between magnetic field $B$ and 
the CPT-Lorentz terms of the kind $\la_{ij} B$. From the experimental point of view, 
it is an interesting situation when the modulus of $\vec{B}$ is sufficiently high in order to compensate 
the external fields effects. 

In Ref. \cite{vit}, for example, 
the idea of the possibility of detecting gravitational waves with atom interferometers 
is explained. In the present work we deal with an analog measuring problem, that is, the weakness 
of the external interaction terms. Our main result (\ref{princ}) shows that the such weakness may 
be compensated by the magnetic field. We intend to perform the 
derivation of the same equations for a gas of spin in a near future. 

\section*{Acknowledgments}
\hspace{0.65cm}The authors wish to thank Prof. Ilya L. Shapiro for the initial discussions about the problem. 
    
BG and MJ are grateful to Funda\c c\~ ao Nacional de Desenvolvimento 
da Educa\c c\~ ao (FNDE) for financial support. Posthumously, BR and BG thank deeply Prof. Wilson Oliveira 
by advices and discussions, during their academic formation.

\renewcommand{\baselinestretch}{0.9}

\end{document}